\documentclass[prx,twocolumn,superscriptaddress,showpacs,floatfix,longbibliography]{revtex4-1}%

\usepackage{amsmath,amssymb}
\usepackage{soul}

\usepackage{lmodern,adjustbox,booktabs,multirow}
\usepackage[para,online,flushleft]{threeparttable}

\usepackage{siunitx}
\usepackage{graphicx}
\usepackage{dcolumn}
\usepackage{bm}
\usepackage{gensymb}
\usepackage{xcolor}
\usepackage{url}
\usepackage{notes2bib}
\DeclareUnicodeCharacter{2212}{-}
\usepackage[colorlinks=true,citecolor=blue]{hyperref}
\usepackage{braket}
\usepackage{afterpage}
\usepackage{booktabs}
\DeclareUnicodeCharacter{2212}{-}

\begin{document}


\title{Ligand-mediated Origin of Altermagnetic Spin-Splitting}

\author{Luigi Camerano}
\email{email: luigi\textunderscore camerano@outlook.it}
\affiliation{Department of Physical and Chemical Sciences, University of L'Aquila, Via Vetoio, 67100 L'Aquila, Italy}
\affiliation{CNR-SPIN L'Aquila, Via Vetoio, 67100 L'Aquila, Italy}

\author{Federico Bisti}
\affiliation{Department of Physical and Chemical Sciences, University of L'Aquila, Via Vetoio, 67100 L'Aquila, Italy}

\author{Gianni Profeta}
\affiliation{Department of Physical and Chemical Sciences, University of L'Aquila, Via Vetoio, 67100 L'Aquila, Italy}
\affiliation{CNR-SPIN L'Aquila, Via Vetoio, 67100 L'Aquila, Italy}

\begin{abstract}
Altermagnets host spin-split electronic bands despite zero net magnetization, opening new routes for spintronics beyond conventional ferromagnets. Going beyond symmetry-based classifications, which specify allowed terms but not their hierarchy, here we use first-principles calculations and Wannier Hamiltonian engineering to uncover the microscopic bonding contributions of altermagnetic spin splitting in the $g$-wave altermagnet Co$_{1/4}$NbSe$_2$. We show that the splitting is captured by a short-range tight-binding model, establishing its local origin. By selectively controlling hopping channels, we demonstrate that the dominant contribution arises not from direct magnetic-ion hopping, but from ligand-mediated hybridization that transfers anisotropy to itinerant states. This identifies ligand-assisted coupling as a dominant mechanism driving altermagnetic spin splitting in intercalated metals, providing a microscopic bridge between minimal models and material-specific electronic structure calculations in altermagnets.
\end{abstract}

\maketitle

Altermagnetism has garnered intense interest as a new class of collinear magnets that combine zero net magnetization with momentum-dependent spin-split electronic bands \cite{smejkal2020,vsmejkal2022_1, vsmejkal2022_2, Radaelli2024, liu2022, Song2025, Ma2021, Hu2025}. Unlike conventional ferromagnets, they retain spin-polarized band structures through the interplay of crystal symmetry and magnetic order \cite{liu2022, Chen2024, jiang2024,Cheong2024}. This unique feature makes them promising for spintronics, enabling spin functionality without stray fields or magnetic interference, with potential applications in spin-current generation, anomalous transport, and ultrafast memory and logic devices \cite{Hernandez2021, Shao2021, Guo2024, Guo2025, Zarzuela20}. 

This growing interest has stimulated an extensive search for realistic altermagnetic materials, combining first-principles predictions with experimental characterization to identify robust platforms for spin-split electronic structure \cite{krempasky2024, Osumi2024,Zhu2024, Amin2024}. Candidate systems now span metals \cite{Jiang2025, Zhang2025, Sakhya2025, Dale2024, DeVita2025}, semimetals \cite{Yang2025, Lu2025}, oxides \cite{Miina2025, Cui2026, Zhang2026, Bandyopadhyay2025}, and layered compounds \cite{Camerano2025, cuiju2026, zhu2025two, sodequist2024, LiuYichen2024}. 
\begin{figure}[!t]
	\centering
	\includegraphics[width=0.95\columnwidth]{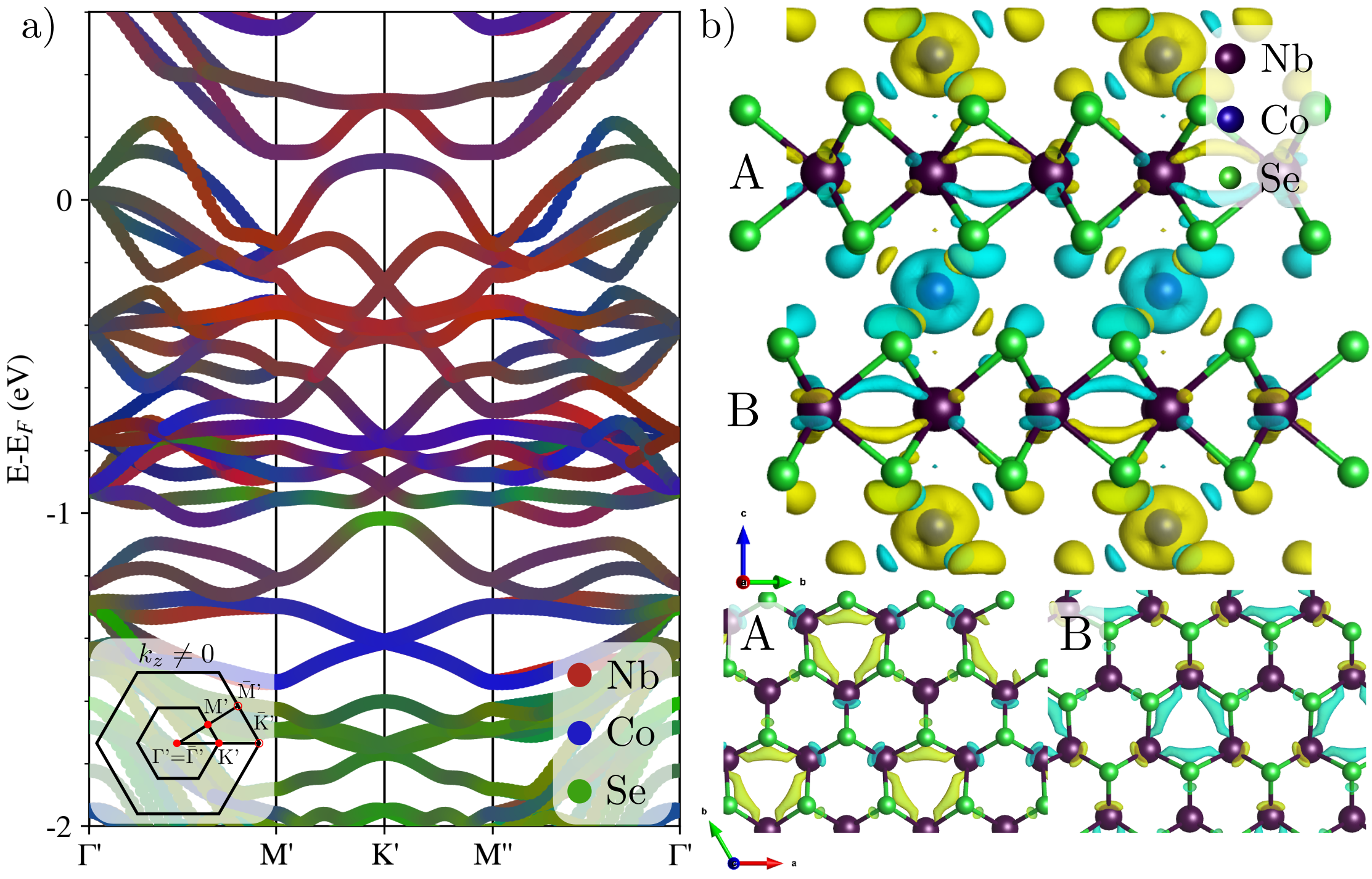}
    \caption{a) Atom-resolved projected band structure of Co$_{1/4}$NbSe$_2$ along high-symmetry lines for $k_z \neq 0$ from DFT calculations. The top-left inset shows the Brillouin Zone (BZ) of the system: primed labels denote points at $k_z \neq 0$, while barred labels refer to the NbSe$_2$ BZ. b) Top panel: magnetization-density isosurfaces with the $b$ axis aligned along the $x$ direction. The labels A and B identify the two inequivalent 2H-NbSe$_2$ layers in the unit cell. Bottom panel: top views of layers A and B, highlighting the induced magnetization on the Nb atoms.}
	\label{fig_1}
\end{figure}
In parallel, minimal Hamiltonian approaches serve as an important theoretical framework, providing a symmetry-based conceptual foundation for understanding altermagnetism. In particular, tight-binding (TB) models on square, honeycomb, Lieb and related lattices \cite{Roig2024,hernandez2025, durrnagel2025, Durnnagel2025, Issing2026, leeb2024spontaneous, Das2024, Kravchuk2025,Camerano2025,jana2025,jana2026} have shown how compensated magnetic sublattices, orbital anisotropy, and crystal symmetry can generate momentum-dependent spin splitting in the absence of both net magnetization and relativistic spin-orbit coupling. 
Together, these studies have opened the way to a classification of altermagnetic states in terms of ferroically ordered magnetic octupoles \cite{spaldin2024,han2025} and symmetry-harmonic decompositions of the spin density \cite{Ubiergo2025}, while identifying characteristic transport responses and providing a unified framework for momentum-dependent spin textures.
\begin{figure*}[!t]
	\centering
	\includegraphics[width=1.95\columnwidth]{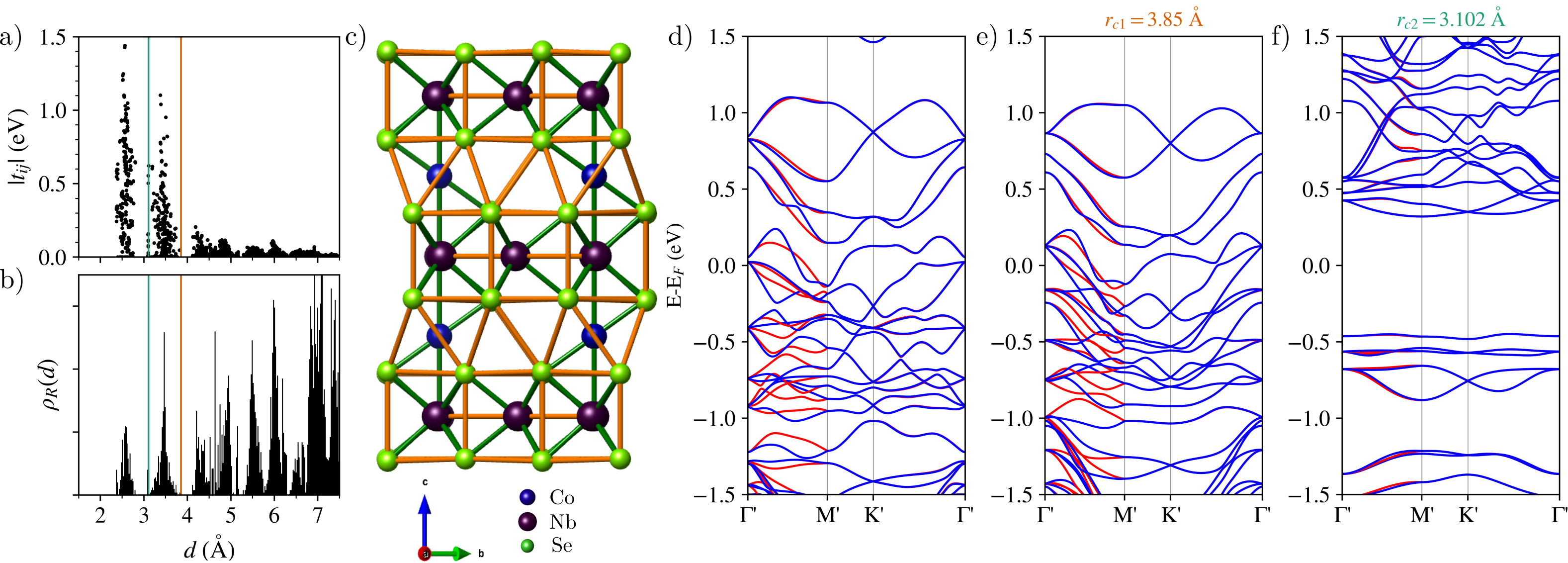}
    \caption{ a) Absolute value of the hopping amplitudes as a function of interatomic distance. b) Histogram of the hopping amplitudes at a given distance. The orange and green lines mark the cutoff radii used in panels e) and f) ($r_{c1}=3.85$ Å and $r_{c2}=3.102$ Å, respectively). c) Unit cell of Co$_{1/4}$NbSe$_2$, where the bond colors indicate the hopping cutoff radius considered. d) Wannier-interpolated band structure along high-symmetry lines, where blue and red denote spin-up and spin-down channels, respectively. Panels e) and f) show the band structures obtained by retaining only hoppings between atoms within cutoff distances $r_{c1}=3.85$ Å and $r_{c2}=3.102$ Å, respectively.}
	\label{fig_2}
\end{figure*}
However, most existing models rely on hopping parameters chosen to satisfy symmetry constraints to capture the existence of altermagnetism, without identifying its microscopic origin in real compounds. In particular, the specific real-space electronic hopping processes, orbital pathways, and chemical bonds determining the altermagnetic spin splitting in real materials remain largely overlooked.
In this paper, we propose a computational framework to uncover the chemical origin of altermagnetism through a real-space decomposition of hopping processes, enabled by first-principles Wannier Hamiltonian engineering. Similar Wannier-based hopping engineering approaches have recently been used to analyze magnetic exchange interactions in insulating 2D magnets \cite{Sabani2025}, here we extend this real-space framework to metallic altermagnets to isolate how momentum-dependent spin splitting is transferred to itinerant host states. As a representative case, we focus on an intercalated transition metal dichalcogenide (TMD) Co$_{1/4}$NbSe$_2$, a system in which chemical composition and symmetry, and consequently the altermagnetic splitting, can be effectively tuned via straightforward processes such as intercalation.

Intercalated layered TMDs with transition metals atoms like Fe, Ni, Co, Mn provide a particularly clean platform to address this problem. Indeed, in the TMD family, strong in-plane itinerancy coexists with strong sensitivity of the electronic and magnetic structure to interlayer stacking symmetry \cite{Camerano2025dark, Li2026, Dale2024, camerano2026} resulting in a variety of electronic and magnetic phases.
For example, in the 2H polytype, the interlayer symmetry combined with the A-type antiferromagnetic order of the intercalants, i.e. ferromagnetic layers stacked antiferromagnetically, enforces $g$-wave altermagnetic spin splitting as demonstrated by recent neutron diffraction \cite{Regmi2025}, muon spin rotation \cite{Graham2025} and angle-resolved photoemission spectroscopy (ARPES) \cite{DeVita2025,Dale2024,Sakhya2025,Sprague2025}. This realizes a layered altermagnetic state in which the altermagnetic symmetry originates from localized moments of the intercalant combined with the symmetry of the host TMD, resulting in the altermagnetic splitting of low-energy TMD derived bands.  
These systems thus provide direct access to the real-space mechanisms that transfer symmetry-driven spin splitting from the intercalant atom to the TMD layer. In particular, they enable the disentanglement of distinct hopping pathways, including direct magnetic–magnetic, ligand-assisted, and host-mediated contributions. \\
Taking Co$_{1/4}$NbSe$_2$ as a model system, we use first-principles calculations and Wannier-based Hamiltonians to identify the hopping channels responsible for transferring altermagnetic spin splitting to itinerant states, thereby providing a real-space framework to analyze the microscopic origin of spin splitting in intercalated metallic TMDs. \\ 
\textit{First-principle electronic structure.} We start by analyzing the electronic and magnetic structure of Co$_{1/4}$NbSe$2$ using first-principles density functional theory (DFT) calculations (see Methods in Appendix). Since our focus is on the non-relativistic altermagnetic spin splitting experimentally observed in Co${1/4}$NbSe$_2$ and related intercalated TMDs \cite{Dale2024, DeVita2025, Sakhya2025, Sprague2025}, all DFT and Wannier-based tight-binding calculations are performed without spin-orbit coupling, thereby isolating the spin splitting arising from exchange interactions and crystal symmetry. As shown in Fig.~\ref{fig_1}a, the bands crossing the Fermi level are predominantly Nb-derived, demonstrating that the low-energy sector is largely itinerant and hosted by the NbSe$_2$ layers. By contrast, Co states are concentrated at higher binding energy around $\sim -1$ eV, while the Se $p$ manifold lies deeper near $\sim -1.5$ eV. This separation of energy scales naturally distinguishes localized magnetic degrees of freedom from itinerant carriers.
\begin{figure*}[!t]
	\centering
	\includegraphics[width=1.99\columnwidth]{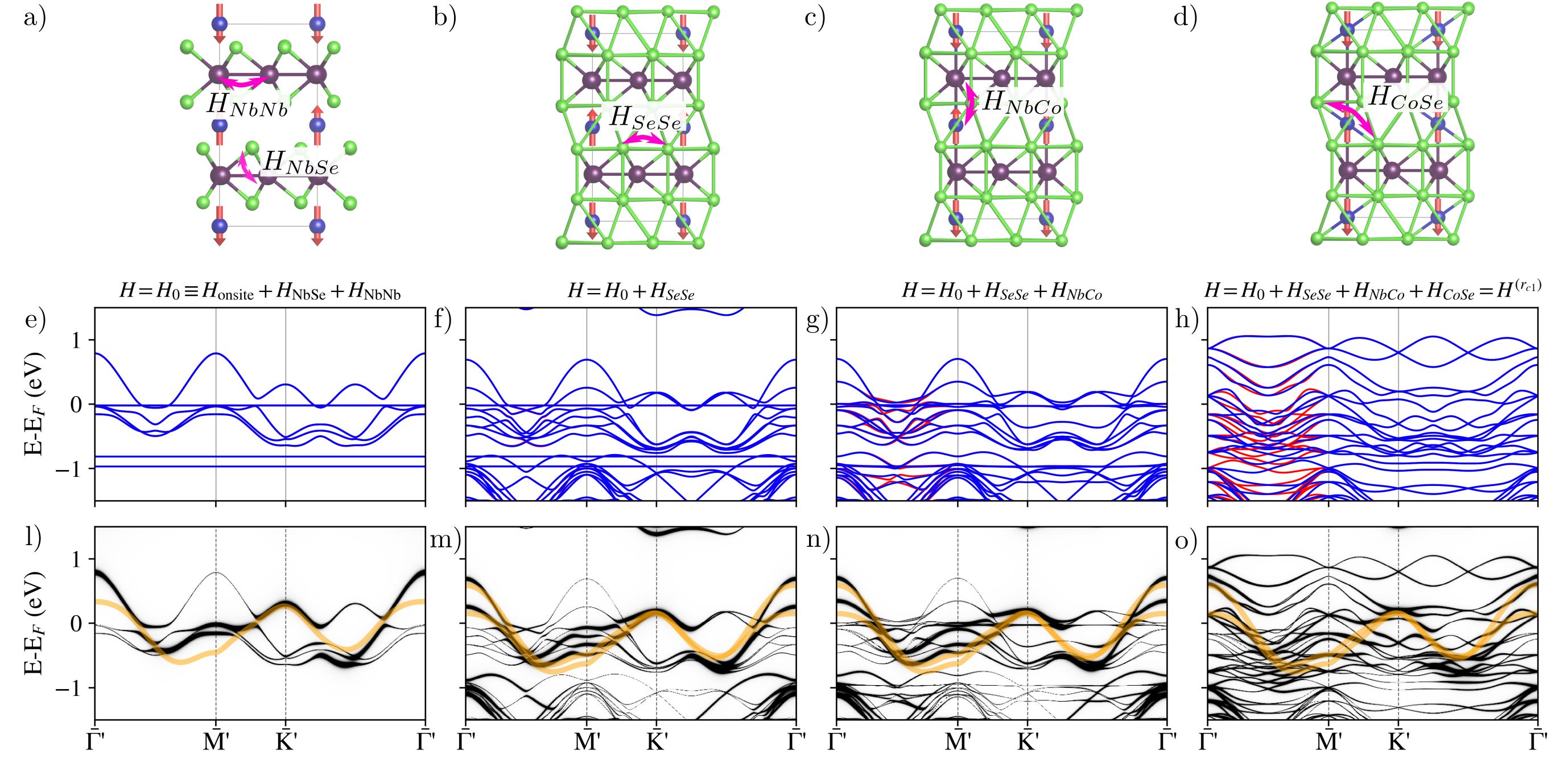}
    \caption{a–d) Unit cell, where only the retained hoppings are indicated as bonds. e-h) Wannier-interpolated band structures obtained by selectively excluding specific hopping terms: e) excluding Co–Se, Se–Se, and Co–Nb hoppings; f) excluding Co–Se and Co–Nb hoppings; g) excluding only Co–Se hoppings; and h) including all hoppings within the cutoff radius $r_{c1}=3.85$ Å. The insets show the unit cell, where only the retained hoppings are indicated as bonds. l–o) Unfolded spectral weight projected onto the NbSe$_2$ unit cell for the same cases shown in panels e–h. In panel l), the solid orange lines correspond to the DFT band structure of a NbSe$_2$ monolayer. In panels m–o), the solid orange lines represent the band structure of 2H-NbSe$_2$.}
	\label{fig_3}
\end{figure*}
The corresponding magnetization density Fig.~\ref{fig_1}b reveals pronounced local moments on Co together with a finite induced polarization on the otherwise nonmagnetic Nb atoms. Importantly, the sign and magnitude of the induced Nb magnetization depend on the inequivalent NbSe$_2$ layers, reflecting the layered A-type antiferromagnetic phase  of the intercalants combined with the symmetry of the 2H phase. The Co magnetic moment is $ \sim 1.4 \: \mu_B$, resulting from the substantial screening by itinerant electrons, as observed in other intercalated compounds \cite{Park2023, camerano2026, Takagi2023}. Thus, in   Co$_{1/4}$NbSe$_2$,  the altermagnetic spin polarization onto the metallic Nb bands originates from the  peculiar coupling between localized Co-moments and itinerant host Nb bands. \\ \textit{Local origin of altermagnetic spin splitting.} To uncover how this is microscopically realized,we construct a Wannier-based tight-binding Hamiltonian from the DFT electronic structure and use it to systematically control the real-space hopping processes in the system, selectively enabling individual hopping channels and orbital pathways.
As long as this approach preserves the overall features of the DFT band structure, it enables a clear disentanglement of the microscopic contributions to the spin splitting. To guide the selection of physically relevant parameters, we first analyze the spatial range of the electronic processes by examining the hopping amplitudes as a function of interatomic distance (Fig. \ref{fig_2}a), along with the corresponding histogram (Fig. \ref{fig_2}b). We then progressively truncate the Wannier Hamiltonian in real space using cutoff distances $r_{c1}$ and $r_{c2}$, indicated in Fig. \ref{fig_2}a–b in orange and green, respectively. The corresponding bonding networks are illustrated in Fig. \ref{fig_2}c, where green bonds denote $r_{c2}$, and both green and orange bonds represent $r_{c1}$.
In Fig. \ref{fig_2}d, we present the Wannier interpolated bands from DFT calculations.

Retaining only the dominant short-range hoppings within $r_{c1}=3.85$~\AA{} (orange and green bonds in Fig.~\ref{fig_2}c) is already sufficient to reproduce the main band dispersions, along with the characteristic altermagnetic spin splitting of the full system (Fig.~\ref{fig_2}d). This indicates that the essential mechanisms are predominantly local and do not rely on longer-range hopping processes. By contrast, further reducing the cutoff to $r_{c2}=3.102$~\AA{}, which eliminates the Nb–Nb and Se-Se hopping network (orange bonds in Fig.~\ref{fig_2}c), substantially modifies the spectrum, driving the system toward an insulating state with only a weak residual spin splitting.
These results show that capturing the overall band structure of Co$_{1/4}$NbSe$_2$ requires the inclusion of Nb–Nb and Se-Se network and  hoppings (orange bonds in Fig.~\ref{fig_2}c), which sustain the dominant altermagnetic response. On the other hand, to clarify the microscopic origin of the spin anisotropy, we performed a symmetry analysis of the Co–Se sublattice. Remarkably, upon removing the Nb atoms while retaining the Co and Se atoms, the system preserves the same spin space group as the full structure, $P\bar{1}6_3/\bar{1}m1\bar{1}m1c^{\infty_{001}}m1$, which allows the observed $g$-wave spin splitting. This demonstrates that Nb is not required to establish the symmetry of the altermagnetic spin texture. Instead, the localized Co moments imprint their nontrivial magnetic symmetry onto the surrounding Se ligands, as directly reflected by the opposite magnetization-density patterns around the spin-up and spin-down Co sites shown in Fig. 1b. The Se-$p$ orbitals therefore provide the symmetry-carrying channel through which the Co magnetic texture is transferred to the itinerant Nb-derived states via hybridization , giving rise to the observed $g$-wave spin splitting (see also SI for band structure without hoppings to Nb ions). \\ 
\textit{Hopping-resolved analysis of the spin splitting.} 

Motivated by the central role of the Nb and Se networks in determining the electronic structure, we take the NbSe$_2$-like electronic structure as a reference manifold and selectively restore individual couplings with the magnetic intercalants. This procedure allows us to isolate how distinct real-space pathways like direct magnetic-ion hopping, ligand-assisted hybridization, and host-mediated processes, contribute to the emergence of altermagnetic spin splitting.  
\begin{figure*}[!t]
	\centering
	\includegraphics[width=1.99\columnwidth]{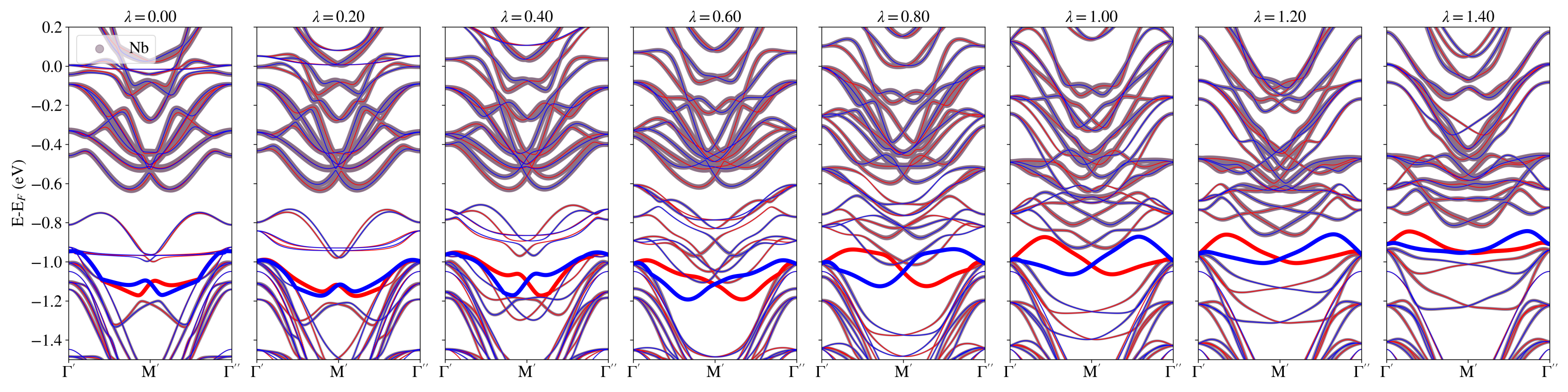}
    \caption{Spin-resolved band structure as a function of the strength of the Co–Se hopping parameter $\lambda \in [0,1.4]$. The size of the markers represents the projection weight on Nb atoms. The thicker bands highlight two representative states, chosen to track the evolution of the spin splitting.}
	\label{fig_4}
\end{figure*}
To this end, we decompose the truncated short-range Hamiltonian as
\begin{equation}
    H^{(r_{c1})}=H_{0}+\sum_{XY} H_{XY},
    \label{eq1}
\end{equation}
Here $H_0$ is the Hamiltonian matrix containing the minimal hoppings terms which are needed to reproduce the  NbSe$_2$ bands close to the Fermi level: hoppings between Nb-Se, and Nb-Nb.
$H_{XY}$ is the part of the Hamiltonian matrix which includes the remaining interactions between  Nb-Co, Co-Se and Se-Se within the cutoff radius $r_{c1}$. 
Starting from the NbSe$_2$ host sector $H_0$, we then selectively activate individual couplings to determine their role in generating the altermagnetic band splitting (see Fig. \ref{fig_3}a-d). 
In Fig.~\ref{fig_3} we report the spin-resolved band structures along the high-symmetry lines of the 2H-NbSe$_2$ BZ at $k_z=0.04\pi$ \AA$^{-1}$, including different interactions (see below) together with its corresponding unfolding weights onto the primitive 2H-NbSe$_2$ unit cell (lower panels). Fig. \ref{fig_3}a-d schematically indicate the Co$_{1/4}$NbSe$_2$ bonds retained in each calculation according to Eq.~(\ref{eq1}). 
Figure~\ref{fig_3}a shows the spectrum obtained from $H_0$ alone, where, as expected, no spin splitting is present. The flat bands present at -1 and 0 are the isolated onsite energies of the Co sublattices. 
Interestingly, we can demonstrate that these bands strongly resembles the DFT bands of the monolayer 1H-NbSe$_2$, after a rigid shift by $0.3$ eV to account for charge transfer from the Co intercalants.
We plot the DFT bands of the 1H phase (defined in the $1\times1$  primitive unit cell) in Fig.\ref{fig_3}l (solid orange bands) compared with the corresponding TB bands (calculated in the $2\times2$ supercell) unfolded in the $1\times1$ unit cell (black lines, with line thickness proportional to the spectral weight).
The differences between these two bands highlight the effect of the intercalant superperiodicity, which reconstructs the host electronic structure through gap openings at folded-band crossings. The strongest reconstruction occurs along $\bar{\Gamma}'\bar{\mathrm{K}}'$, with smaller hybridization mini-gaps visible along $\bar{\Gamma}'\bar{\mathrm{M}}'$ and $\bar{\mathrm{K}}'\bar{\mathrm{M}}'$.

We then selectively include $H_{\mathrm{SeSe}}$, with the resulting spectrum shown in Fig.~\ref{fig_3}f. The Se-Se hopping network restores the three-dimensional character of the electronic structure while preserving zero spin splitting, since couplings to the magnetic Co atoms are still absent. This is further illustrated in Fig.~\ref{fig_3}m, where comparison with the DFT 2H-NbSe$_2$ bands shows that the splitting at $\bar{\Gamma}'$ and $\mathrm{M}'$ originates from interlayer hybridization \cite{Weber2018,Camerano2025}, while the locations of the gap openings remain essentially unchanged from the $H_0$ case. In addition, Se-derived down-dispersive bands emerge below $-1$ eV, consistent with the projected DFT spectrum of Fig.~\ref{fig_1}a. The further inclusion of $H_{\mathrm{NbCo}}$, shown in Fig.~\ref{fig_3}g-n, only weakly modifies the overall band structure but has a qualitative impact on the spin polarization. Remarkably, although $H_{\mathrm{NbCo}}$ does not itself contain anisotropic spin-dependent hoppings between antiferromagnetically ordered Co sites typically required to generate $g$-wave altermagnetism, it nevertheless induces a finite spin splitting through its indirect coupling to the anisotropic Se network. In particular, despite the absence of direct anisotropic Nb-Co hopping, the Nb-derived bands near the Fermi level become strongly spin polarized, demonstrating that the altermagnetic splitting is efficiently transferred to the itinerant states via ligand-mediated processes.

Indeed, as shown in the SI, including $H_{\mathrm{NbCo}}$ while neglecting $H_{\mathrm{SeSe}}$ preserves the band degeneracy. This demonstrates that the spin splitting originates from an indirect ligand-mediated pathway involving the Se network.
As seen in Fig.~\ref{fig_3}n, the comparison with the 2H-NbSe$_2$ bands remains meaningful at this stage. In particular, the folded bands along $\bar{\Gamma}'\bar{\mathrm{K}}'$ acquire finite spectral weight, partially filling the gap opened by the superperiodicity, while retaining the overall structure of the host-derived states. 
Finally, restoring $H_{\mathrm{CoSe}}$ completes the full Hamiltonian $H^{(r_{c1})}$ and fully recovers the $g$-wave altermagnetic spin splitting (see Fig.~\ref{fig_3}h). In contrast to $H_{\mathrm{NbCo}}$, this term directly introduces bond anisotropy tied to the Co sublattice, leading to a pronounced reconstruction of the electronic structure. As shown in Fig.~\ref{fig_3}o, bands are strongly split and additional mini-gaps develop along $\bar{\Gamma}'\bar{\mathrm{K}}'$, highlighting the central role of the intercalant in shaping the low-energy spectrum. The previously flat Co-derived states acquire a finite in-plane dispersion, signaling enhanced hybridization with the host.
This analysis reveals a clear hierarchy of microscopic processes underlying altermagnetic spin splitting. While the NbSe$_2$ host sector sets the itinerant electronic structure and the Se network establishes the three-dimensional connectivity, neither is sufficient to generate spin splitting. The inclusion of $H_{\mathrm{NbCo}}$ introduces a finite but indirect contribution mediated by the ligand network, whereas the dominant mechanism arises from $H_{\mathrm{CoSe}}$, which transfers the symmetry-breaking anisotropy of the Co sublattice to the itinerant Nb bands. These results identify ligand-assisted hybridization as the key channel through which altermagnetic spin splitting emerges in the metallic state. \\ \textit{Tuning ligand-mediated spin splitting.} To quantify this mechanism, we continuously tune the Co–Se hybridization strength via a scaling dimensionless parameter $\lambda$, such that $H_{\mathrm{CoSe}} \rightarrow \lambda H_{\mathrm{CoSe}}$. Fig.~\ref{fig_4} shows the evolution of the band structure as a function of $\lambda$ along the $\Gamma'M'\Gamma''$ high-symmetry line, with point size indicating the Nb orbital character. For $\lambda=0$, the system exhibits weak spin splitting. Upon increasing $\lambda$, a momentum-dependent spin polarization develops in both Co- and Se-derived bands (one highlighted by thicker lines) and is subsequently transferred to Nb-derived states near the Fermi level. With increasing $\lambda$, initially flat Co-derived states approach the Nb bands and become progressively hybridized via the Se network. At large $\lambda$, the NbSe$_2$ band structure is strongly renormalized and the spin splitting at the Fermi level is reduced. In contrast, small deviations around $\lambda=1$ have only a minor effect on the altermagnetic spin splitting, in agreement with the weak material dependence of the spin-splitting observed across different TMD hosts and intercalant species \cite{Roberts2026}. This weak sensitivity reflects that variations in material-specific parameters primarily renormalize microscopic Hamiltonian details without altering the overall scale of the induced spin splitting. This smooth evolution demonstrates that the altermagnetic spin splitting is continuously transmitted to the itinerant sector through ligand-mediated hybridization, providing direct microscopic evidence of its origin. \\ 
The intercalant-ligand bonding underlying this mechanism is experimentally accessible through core-level spectroscopy. In Co$_{1/4}$NbSe$_2$ and related intercalated TMDs, distinct ligand core-level components reveal the different electronic environments of surface ligands with reduced intercalant coordination and ligands retaining intercalant bonds \cite{DeVita2025,Camerano2026_plasmon}. These spectroscopic fingerprints support the importance of intercalant-ligand hybridization and motivate further investigation of how ligand coordination and bonding influence the spin splitting of itinerant states. \\
\textit{Conclusions.} In this work, we have identified the microscopic origin of altermagnetic spin splitting in Co$_{1/4}$NbSe$_2$ by combining first-principles calculations with a Wannier-based real-space decomposition of the electronic Hamiltonian. By selectively resolving and tuning individual hopping channels, we disentangled the different contribution to the altermagnetic spin splitting demonstrating that it is not rooted in direct magnetic-ion hopping, but is instead primarily generated through ligand-mediated Co–Se–Nb pathways. These processes transfer the symmetry of the magnetic sublattice to itinerant metallic states, establishing a direct real-space mechanism for altermagnetic band splitting in metallic systems. By connecting symmetry-based tight-binding descriptions with material-specific modeling, this work offers a clear real-space picture of how ligand hybridization shapes spin splitting in intercalated layered conductors, highlighting the importance of ligand physics.

\section{Appendix}
\textbf{Methods.} Density functional theory calculations were performed using the Vienna ab-initio Simulation Package (VASP) \cite{Kresse:1993,Kresse:1999}, using the generalized gradient approximation (GGA) in the Perdew-Burke-Ernzerhof (PBE) parametrization for the exchange-correlation functional \cite{Perdew:1996}. Interactions between electrons and nuclei were described using the projector-augmented wave method. Energy thresholds for the self-consistent calculation was set to $10^{-6}$ eV and force threshold for geometry optimization $10^{-4}$ eV $\text{Å}^{-1}$. A plane-wave kinetic energy cutoff of 550 eV was employed. The Brillouin zone was sampled using an 6 \texttimes{} 6 \texttimes{} 4 Gamma-centered Monkhorst-Pack grid. We used experimental lattice parameter $a= 6.9178$ \AA \: and $c= 12.390$ \AA \: as determined by neutron diffraction \cite{Regmi2025}. All calculations were performed without an additional Hubbard $U$ correction. This choice is motivated by recent ARPES studies showing good agreement between the measured electronic structure, including the low-energy band dispersions and altermagnetic spin splitting, and calculations performed without a Hubbard correction \cite{DeVita2025}. \\ Maximally localized Wannier functions \cite{wannierMLWF1997,wannierMLWF2001,wannierMLWF2012} are obtained by using Wannier90 code \cite{Pizzi2020} using atomic projection on Co-$d$, Nb-$d$ and Se-$p$ orbital. The real-space Hamiltonian obtained from DFT bands is analyzed using the open-source code PythTB \cite{Cole_Python_Tight_Binding_2025}.

\section{Data availability statement}
All data that support the findings of this study are included within the article and supplementary materials.

\renewcommand{\thefigure}{A\arabic{figure}}
\setcounter{figure}{0}


\section{Acknowledgements} 
 This work was funded by the European Union-NextGenerationEU under the Italian Ministry of University and Research (MUR) National Innovation Ecosystem Grant No. ECS00000041 VITALITY-CUP E13C22001060006.

\newpage

\clearpage 

\bibliography{bibliography.bib}

\end{document}